# Influence of paroxysmal activity on background synchronization in epileptic records

Jesús Pastor[1] and Guillermo Ortega[1]

[1] Instituto de Investigación Sanitaria Hospital de la Princesa, Madrid

**Abstract**

The presence of spikes and sharp waves in the recordings of epileptic patients may contaminate background signal synchronization in different ways. In this Technical Note, we present a simple procedure for assessing whether a particular synchronization method should be used (or not) with data from neurophysiological recordings commonly used to evaluate epilepsy. The information provided by this procedure makes it possible to differentiate true background synchronization from spike synchronization. This issue is particularly relevant when differentiating between the mechanisms underlying the onset of interictal epileptiform discharges and limbic network dynamics.

*Abbreviations*: Interictal Epileptiform Discharges (IED); Electroencephalogram (EEG); Electrocorticography (ECoG); Foramen Ovale Electrodes (FOE); Depth Electrodes (DE); Temporal Lobe Epilepsy (TLE); Phase Synchronization (PS); Mutual Information (MI); Measure of Synchronization (MoS); Influence of IED over Background Activity (IoB).

The presence of interictal epileptiform discharges (IED) in neurophysiological recordings enables us to differentiate between epileptic and nonepileptic patients[1]. Correct identification of IED requires simultaneous occurrence of discharges in at least 2 neighboring contacts caused by the physiological field of the generator. However, co-occurrence of IED in distant electrodes in epileptic patients is usually assessed as true synchronized activity between the pathophysiological structures involved, whether during the interictal[2] or preictal periods[3].

This finding contrasts with a more recent approach, which uses the full interictal background signal to assess synchronization[4]. Thus, the relationship between interictal background signal and IED content of the signal is a key issue that was first addressed in Bettus et al.[5] and explored in Ortega et al.[6]. In this Technical Note, we investigate how and to what extent the presence of IED influences signal synchronization in typical neurophysiological recordings of epileptic patients.

Several neurophysiological techniques—EEG, foramen ovale electrodes (FOE), electrocorticography (ECoG), and depth electrodes (DE)—are routinely used to lateralize/localize epileptogenic areas in patients with drug-resistant temporal lobe epilepsy (TLE). Identification of the epileptogenic zone is the major goal of the neurophysiologist. However, it is essential to associate this zone with other important areas, such as the ictal onset zone and the irritative zone, where IED actually appear. The duration of an IED[7] on an EEG is <200 ms. Sharp waves have a duration of 70-200 ms, whereas that of spikes is <70 ms. Although duration is similar in different recording techniques, the quantity of IED present may vary largely from one recording technique to another. A typical scalp EEG has an average spike frequency of 1 spike/min (60 spikes/h), although this value can increase to 4 spikes/min or more. In an ECoG recording, these values should be multiplied by 10, as Tao et al.[8] suggested. Table I summarizes typical values of spike frequencies found in the literature[2,6,8,9] for each recording technique. Given that a typical spike lasts at most 70 ms and a sharp wave 200 ms, we can estimate the "IED content" (i.e., the percentage of time that IED occupies in the background signal) using the method set out below. In order to adopt a conservative approach, we set the duration of IED at 200 ms or 0.2 seconds. A spike frequency of 4 spikes/min (in the case of EEG) is equivalent to (0.2 s x 4 spikes)/60 s = 0.013 of "spike content" in the recording (i.e., an IED content of 1.3%).

By following the same line of reasoning for other values in Table I, we obtain the IED content in the background signal (see column 4 of Table I).

Figure 1A shows a representative FOE recording from a patient with right TLE. Four recordings (2 from the left side and 2 from the right side) show mesial interictal activity. Activity from right contacts, however, displays IED in at least 4 locations (rectangles). Calculation of synchronization between both right FOE contacts based on the Pearson correlation coefficient yields a value of 0.87. When IED activity is eliminated from the recording, the value of the correlation drops to 0.80 because of the strong synchronization between these high-amplitude IED. Thus, synchronization increases by almost 9% as a result of the presence of IED. The Pearson correlation between the left FOEs is 0.93.

In order to address the question of whether paroxysmal activity "contaminates" synchronization estimates, we implemented the following procedure. Several simulated IED, each one represented by a single sine wave cycle with period equal to 200 data points, were inserted simultaneously into 2 correlated white Gaussian signals in such a way that they occupied a specific percentage of the recording. Figure 1B shows three simulated IED occupying 10% of the whole recording (200x3 data points of three simulated IED in a record of 6,000 data points). The amplitude of the sine wave is 3 times the standard deviation ($\sigma$) of the background signal. Both background signals were generated with a specified value of correlation ($\rho$) between them. Specifically, we generated 2 stochastic signals with a bivariate normal distribution, a given mean value ($\mu$=0 in every case), and a given covariance matrix. In the covariance matrix, we fixed the $\sigma$ of both signals at 1 and changed the correlation so that $0<\rho<1$. Several programming packages (e.g. R) enable the above procedure to be implemented easily[10]. Because IED must be clearly differentiated from background activity[7], we calculated different values of simulated IED amplitude in relation with the $\sigma$ of the background signal. We call this ratio A2S, that is:

$$A2S = \frac{Amplitude(IED)}{\sigma(background\ signal)} \qquad (1)$$

Because $\sigma$ was set at 1 in each run, A2S is always equal to the amplitude of the simulated IED. Three values of A2S were used, 1, 3 and 5. In the example shown in Figure 1B, the correlation value $\rho$ between the background signals is 0.375.

Three frequently used methods were applied to assess synchronization[6,11,12], namely, Pearson correlation, phase synchronization (PS), and mutual information (MI). We refer to these methods generically as measures of synchronization (MoS).

Lastly, we generated 2 correlated Gaussian signals of 60,000 data points each in length (60 seconds), with $\mu$=0, $\sigma$=1, and $0<\rho<1$. Synchronization between both signals was measured using Pearson, PS, and MI. The procedure was repeated using signals containing different percentages of IED. The proportion of simulated IED increased from 0% (no IED) to 100% (300 IEDs).

The influence of IED on background synchronization (IoB) can be quantified as follows:

$$IoB = \frac{MoS(stoch, stoch)}{MoS(stoch + IED, stoch + IED)} \qquad (2)$$

As expected, the presence of simultaneous IED in both signals increases the value of synchronization and, thus, the denominator increases faster than the numerator. In the example of right contacts of Figure 1A, IoB (Pearson)=0.80/0.87 = 0.92.

Figure 2A shows IoB for each MoS (rows) and different values of A2S (columns). IoB close to 1 (white) implies poor influence of IED on the synchronization measured. This is clear in the lower part of each panel, where the percentage of IED is very low and, therefore, both measures are similar. In contrast, IoB close to 0 (red) implies that the MoS in the IED-contaminated signals is much higher than in the case of pure signals. In short, the yellowish part of the graph is the safer area to analyze synchronization between signals, without contamination by IED. The vertical dot-dashed line in Figure 2A corresponds to $\rho=0.5$ between the background signals. Horizontal solid, dotted, dot-dashed, and dashed lines correspond to EEG, FOE, ECoG, and DE maximum IED content (column 4 in Table 1), respectively. The intersection of the vertical line with each horizontal line is shown in Table 1 (columns 5 to 13) and plotted in Figure 2B, for every MoS.

Figure 2A is illustrative in several aspects. First, for high values of the underlying $\rho$, the influence of IED is less significant in the synchronization measured. For Pearson and PS, it seems that poor influence of IED synchronization contaminates signal synchronization above $\rho > 0.5$ (approximately). MI seems to need far greater values of measured synchronization, because, for $\rho=0.8$ (for the case of A2S=5) and 30% IED content, MI(stoch,stoch) is approximately equal to 0.3 times MI(stoch+IED,stoch+IED), which means that IED synchronization contaminates 3.3 times the value of the MI estimate. Contamination is even worse at lower values of correlation.

Second, while EEG, FOE, and ECoG recordings are almost always in safer areas, at least for values of $\rho>0.4$, recordings from DE must be interpreted with caution (see Bourien et al[2]). Moreover, when MI is used as the selected synchronization measure, the influence of IED almost always severely contaminates full signal synchronization. Third, PS is the most robust synchronization method, with high-amplitude IED content, whereas MI performs worse than the other 2 methods, because IED content considerably influences the synchronization measure, even at very low percentages of IED content.

Figure 2A was constructed using basic estimates of synchronization measures, particularly PS and MI. Several methods can be used to improve estimations[9,10] however. One point to be remarked is the use of white noise to model background activity instead of the more appropriate colored noise[13,14] (page 119 an so on in the book of Buzsáki[14] and references therein). Although the correlation structure of the EEG is of extreme importance at the time to model background activity, in the present work white noise has been used because of: a) A simple theoretical expression is available for simulation of correlated Gaussian noise which allows generating easily several correlated signals in a definite range of correlation values and b) white noise time series are highly stationary, as opposed to the case of $1/f^{\alpha}$ noise. Correlation between two non-stationary time series therefore could obscure our principal point.

More realistic simulations, such as those based on truly cortical models with epileptogenic activity and connectivity between different locations[15,16], can be used in a more in-depth perspective. However, such an approach is beyond the scope of this study.

In addition to the above considerations, we think that a procedure such as that shown in Figure 2A would help to explore and evaluate synchronization estimates obtained on actual neurophysiological recordings from epileptic patients. Likewise, it would help in deciding which kind of MoS should be used in each type of recording in order to assess background signal synchronization without contaminating IED synchronization.

*Sources of support*: This work has been funded by grants from Fundación Mutua Madrileña, Instituto de Salud Carlos III, through PS09/02116 and PI10/00160 projects, and PIP Nº 11420100100261 CONICET. GJO is member of CONICET, Argentina.

**References**

1. Noachtar S, Rémi J. The role of EEG in epilepsy: A critical review. Epilepsy & Behavior 2009; 15: 22–33.
2. Bourien J, Bartolomei F, Bellanger JJ, Gavaret M, Chauvel P, Wendling, F. A method to identify reproducible subsets of co-activated structures during interictal spikes. Clinical Neurophysiology 2005; 116: 443–455.
3. Spencer SS, Spencer DD. Entorhinal-Hippocampal Interactions in Medial Temporal Lobe Epilepsy. Epilepsia 1994; 35(4): 21-727.
4. Mormann F, Lehnertz K, David P, Elger CE. Mean phase coherence as a measure for phase synchronization and its application to the EEG of epilepsy patients. Physica D 2000; 144: 358–369.
5. Bettus B, Wendling F, Guye M, Valton L, Régis J, Chauvel P, Bartolomei F. Enhanced EEG functional connectivity in mesial temporal lobe epilepsy. Epilepsy Res 2008; 81: 58-68.
6. Ortega GJ, Herrera Peco I, García de Sola R, Pastor J. Impaired mesial synchronization in temporal lobe epilepsy. Clinical Neurophysiology 2010; 122(6): 1106-1116.
7. Walczak TS, Jayakar P, Mizrahi EM. Interictal Electroencephalography. In: Epilepsy: A Comprehensive Textbook, 2nd Edition. 2008. Engel, J. & Pedley, T.A., Lippincott Williams & Wilkins.
8. Tao JX, Ray A, Hawes–Ebersole S, Ebersole JS. Intracranial EEG substrates of scalp EEG interictal spikes. Epilepsia 2005; 46: 669–676.
9. Clemens Z, Janszky J, Szucs A, Békésy M, Clemens B, Halász P, Interictal epileptic spiking during sleep and wakefulness in mesial temporal lobe epilepsy: a comparative study of scalp and foramen ovale electrodes. Epilepsia 2003; 44(2): 186–92.
10. Genz A, Bretz F, Miwa T, M, X, Leisch F, Scheipl F, Hothorn T. 2011. mvtnorm: Multivariate Normal and t Distributions. R package version 0.9-9991. URL http://CRAN.R-project.org/package=mvtnorm
11. Lehnertz K, Bialonski S, Horstmann MT, Krug D, Rothkegel A, Staniek M, Wagner T. Synchronization phenomena in human epileptic brain networks. Journal of Neuroscience Methods 2009; 183**:** 42–48.
12. Pereda E, Quian Quiroga R, Bhattacharya J. Nonlinear multivariate analysis of neurophysiological signals. Progress in Neurobiology 2005; 77: 1–37.

13. Milstein J, Mormann F, Fried I, Koch C. Neuronal Shot Noise and Brownian $1/f^2$ Behavior in the Local Field Potential. PLoS ONE 2009;4(2): e4338
14. Buzsáki G (2006) Rhythms of the brain. Oxford: Oxford University Press
15. Wendling F, Bellanger JJ, Bartolomei F, Chauvel P. Relevance of nonlinear lumped-parameter models in the analysis of depth-EEG epileptic signals. Biol. Cybern 2000: 83; 367-378.
16. Wendling F, Ansari-Asl K, Bartolomei F, Senhadji L. From EEG signals to brain connectivity: A model-based evaluation of interdependence measures Journal of Neuroscience Methods 2009: 183; 9–18

Table 1 caption: Spike frequency values (average and maxima) for each type of recording. The column "Maximum percentage of time" represents the percentage of the recording corresponding to spikes. Representative values of IoB correspond to correlated Gaussian signals with $\rho=0.5$, for different values of A2S (1, 3, 5). Shaded areas (A2S=3) correspond to the values for the intersection of the vertical line ($\rho=0.5$) with the horizontal lines (EEG, FOE, ECoG, and DE) in Figure 2A.

Figure 1 caption: (A) Interictal signals of FOE recordings from a patient with right TLE. IED are highlighted in the right contact and occupy approximately 15% of the total length of the recording. (B) 2 correlated Gaussian recordings with $\mu=0$, $\sigma=1$, length of 6,000 data points, and a correlation between them of $\rho=0.375$. Identical simulated IED, with A2S=3, are inserted simultaneously in both signals. Simulated IED account for 10% of the total length of the recording. IoB in this case is 0.86. a.u. stands for arbitrary units.

Figure 2 caption: (A) Levelplot of IoB. The MoS is the Pearson correlation in the first row, PS in the second row, and MI in the third row. Representative values are depicted in the middle column (A2S=3). Horizontal lines represent typical percentages of IED times for each recording type (see Table 1). The vertical line, $\rho=0.5$, is used to guide the eye. (B) Bar plot of representative values of IoB (see Table 1).

| Method | Spike frequency | | Maximum percentage of IED content | IoB at $\rho=0.5$ | | | | | | | | |
|---|---|---|---|---|---|---|---|---|---|---|---|---|
| | Average (spikes/min) | Maximum (spikes/min) | | Pearson coefficient | | | Phase synchronization | | | Mutual information | | |
| | | | | 1 | 3 | 5 | 1 | 3 | 5 | 1 | 3 | 5 |
| EEG | 1 | 4 | 1.3 % | 0,99 | 0,96 | 0,90 | 0,99 | 0,98 | 0,99 | 0,98 | 0,91 | 0,89 |
| FOE[6,9] | 3.5 | 30 | 10.0 % | 0,95 | 0,75 | 0,63 | 0,88 | 0,87 | 0,87 | 0,66 | 0,45 | 0,35 |
| ECoG[8] | 10 | 40 | 13.3 % | 0,94 | 0,71 | 0,61 | 0,85 | 0,84 | 0,84 | 0,58 | 0,38 | 0,29 |
| DE[2] | 55.3 | 100 | 33.0 % | 0,84 | 0,59 | 0,54 | 0,68 | 0,67 | 0,68 | 0,28 | 0,19 | 0,14 |

Table 1

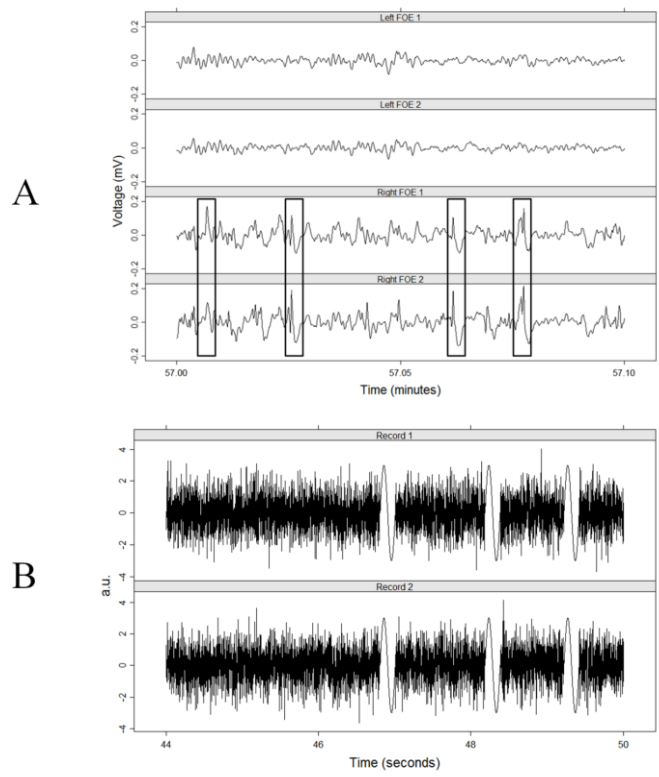

Figure 1

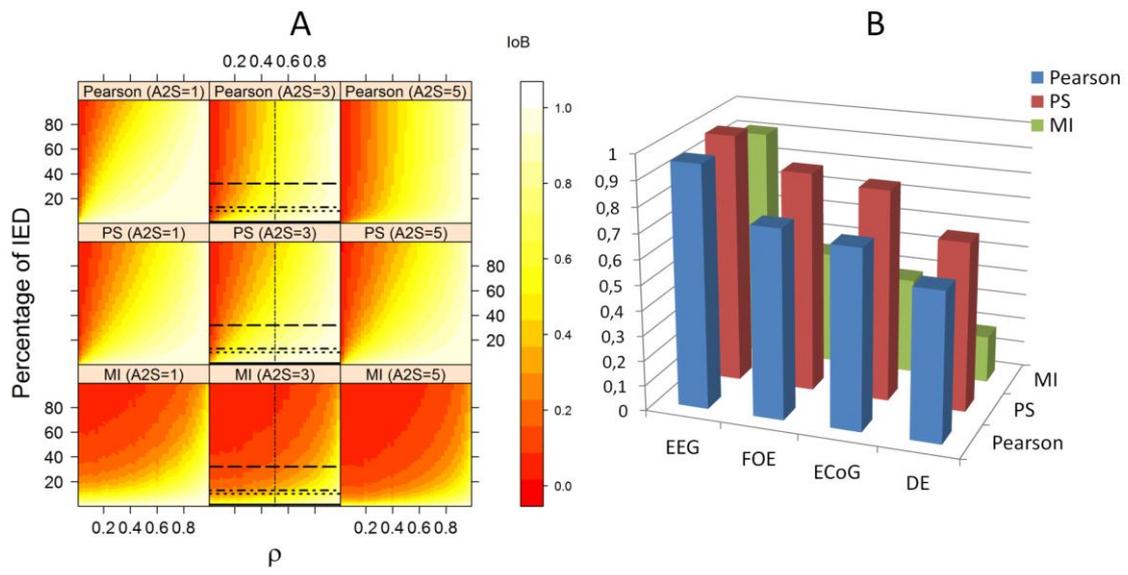

Figure 2